\documentclass[a4paper,fleqn]{cas-dc}

\usepackage[numbers]{natbib}

\usepackage{graphicx}
\usepackage{color}

\def\tsc#1{\csdef{#1}{\textsc{\lowercase{#1}}\xspace}}
\tsc{WGM}
\tsc{QE}

\begin{document}
\let\WriteBookmarks\relax
\def\floatpagepagefraction{1}
\def\textpagefraction{.001}

\shorttitle{Carrier doping of Bi$_2$Se$_3$ surface by
  chemical adsorption -- a DFT study}

\shortauthors{C. Fan, K. Sakamoto, P. Krüger}  

\title [mode = title]{Carrier doping of Bi$_2$Se$_3$ surface by
  chemical adsorption -- a DFT study}



%

\author[1]{Cheng Fan}[]




\credit{<Credit authorship details>}

\affiliation[1]{organization={Graduate School of Science and Engineering, \
Chiba University}, city={Chiba}, postcode={263-8522}, country={Japan}}

\author[2,3,4]{Kazuyuki Sakamoto}[orcid=0000-0001-9507-6435]



\affiliation[2]{organization={Department of Applied Physics, Osaka University}, city={Osaka}, postcode={565-0871},country={Japan}}
\affiliation[3]{organization={Spintronics Research Network Division, OTRI,Osaka University}, city={Osaka}, postcode={565-0871},country={Japan}}
\affiliation[4]{organization={Center for Spintronics Research Network, Osaka University}, city={Osaka}, postcode={565-0871},country={Japan}}

\author[1,5]{Peter Kr\"uger}[orcid=0000-0002-1247-9886]

\cormark[1]


\affiliation[5]{organization={Molecular Chirality Research Center, Chiba University}, city={Chiba}, postcode={263-8522}, country={Japan}}

\cortext[1]{Corresponding author. E-mail: pkruger@chiba-u.jp}



\begin{abstract}
Bi$_2$Se$_3$ is one of the most promising topological insulators, but it suffers from intrinsic n-doping due to Se-vacancies, which shifts the Fermi level into the bulk conduction band, leading to topologically trivial carriers. Recently it was shown that this Fermi-level shift can be compensated by a locally controlled surface p-doping process, through water adsorption and XUV irradiation.
Here, the microscopic mechanism of this surface doping is studied by means of density functional theory (DFT) focusing on the adsorption of H$_2$O, OH, O, C and CH on Bi$_2$Se$_3$. We find that water adsorption has a negligible doping effect while hydroxyl groups lead to n-doping. Carbon adsorption on Se vacancies gives rise to p-doping but it also strongly modifies the electronic band structure around the Dirac point. Only if the Se vacancies are filled
with atomic oxygen, the experimentally observed p-doping without change of the topological surface bands is reproduced. Based on the DFT results, we propose a reaction path where photon absorption gives rise to water splitting and the produced O atoms fill the Se vacancies. Adsorbed OH groups appear as intermediate states and carbon impurities may have a catalytic effect in agreement with experimental observations.
\end{abstract}




\maketitle

\section{Introduction}
In a topological insulator, the surface becomes metallic through
the appearance of edge-states, 
whose band dispersion forms a Dirac cone near the Fermi level.
By virtue of time-inversion symmetry, the edge states are spin-momentum
locked and topologically protected from back-scattering by non-magnetic
impurities~\cite{1,2,3,4}. These properties make topological insulators
a very promising class of materials for low-consumption electronic devices.
Bi$_2$Se$_3$ is one of earliest and best studied
three-dimensional topological insulators~\cite{5,6,7,8,9,10}.
The Bi$_2$Se$_3$ crystal has trigonal symmetry but is more easily
viewed as a hexagonal system, where the unit cell contains 15 atomic
layers, or three quintuple layers (QL), a sequence of five covalently bonded
atomic layers (Se-Bi-Se-Bi-Se). The QLs are held
together by weak van-der-Waals interactions.
This makes Bi$_2$Se$_3$ a layered material with a stable,
Se-terminated (0001) surface.
However, other surface terminations
and reconstructions may be stabilized under special conditions
and lattice defects exist~\cite{10d,10e,10f}.

Bulk Bi$_2$Se$_3$ is a semiconductor
  with a direct band gap of 0.22~eV~\cite{10c}. At the (0001) surface,
topological edge states with a Dirac cone dispersion appear,
such that spin-momentum locked carriers cross the Fermi-level and the surface
becomes conducting. These topological surface states were recently
observed even in amorphous~\cite{10a} and mesoporous Bi$_2$Se$_3$~\cite{10b}
samples, which calls for a thorough investigation of the role of
lattice imperfections.
Many physical properties of Bi$_2$Se$_3$ have been studied using DFT,
including dielectric~\cite{5a,14a} and thermoelectric response~\cite{5b} as well as the interface properties in ferromagnetic In$_2$Se$_3$/Bi$_2$Se$_3$ heterostructures~\cite{5c}. Bi$_2$Se$_3$ is also considered a potential catalyst
for the hydrogen evolution reaction~\cite{12a}.
  In order to tune the electronic properties of Bi$_2$Se$_3$, doping with various elements (Te, Sb, Dy, Au, Ca, transition metals etc) has been
studied both experimentally and theoretically~\cite{15a,15b,15c,15d,15e,15f}.

In general, the Bi$_2$Se$_3$(0001) surface is very stable and chemically
inert, which makes it an attractive material for device applications.
Exposure to water and oxygen at room temperature does
not give rise to irreversible surface reactions~\cite{5}.
However, Bi$_2$Se$_3$ is naturally $n$-doped, due the presence of
Se vacancies, which are the dominant intrinsic defects.
As a consequence, the Fermi level is raised into the conduction band,
which leads to trivial metallic transport in the bulk
and scattering of the topological surface band states with bulk carriers,
i.e. a loss of the topological protected nature of the edge states.
Therefore the intrinsic n-doping by Se vacancies must be compensated by
p-doping if Bi$_2$Se$_3$ is to be used as a topological material
in technological applications.
To this end, bulk and surface doping with various species have been
investigated. Substitution of Ca for Bi~\cite{15e}, as well as
adsorption of Ru, NO$_2$~\cite{6,12,13}, O$_2$~\cite{9,14,15} or O was
found to lead to p-doping and thus may repair
the defect-induced n-doping of Bi$_2$Se$_3$.
However, in these methods the doping level was not shown to be stable
against aging and atmospheric conditions, which is a crucial requirement
for device applications.
Recently, Sakamoto {\it et al.}~\cite{16} have developed a new method of
surface p-doping by water adsorption and UV or X-ray irradiation.
This method is stable under atmospheric conditions.
Importantly, through radiation,
the doped surface area can be controlled spatially at the nanometer scale,
opening new opportunities for device applications.

Here we present a density functional theory (DFT) study on the adsorption of H$_2$O, OH,
O, C and CH on Bi$_2$Se$_3$ surface and examine the
doping effect in the light of the experimental results of Ref.~\cite{16}.
We find that H$_2$O physisorbs on the surface with negligible
doping effect, whereas all other species lead to chemisorption which strongly
affects the surface band structure. We compare adsorption on 
the pristine and defective surface and find that adsorption at
the Se vacancy is energetically favored for O and OH but not for C.
The experimentally observed p-doping with a well preserved Dirac cone can
only be reproduced by O atom adsorption at a Se vacancy.
This strongly suggests that the photo-induced p-doping
process of Ref.~\cite{16} involves water splitting. Based on the
DFT results we propose a mechanism for this surface reaction,
where carbon impurities may act as a catalyst.

\begin{figure}
  \begin{center}
   \includegraphics[width=0.7\columnwidth]{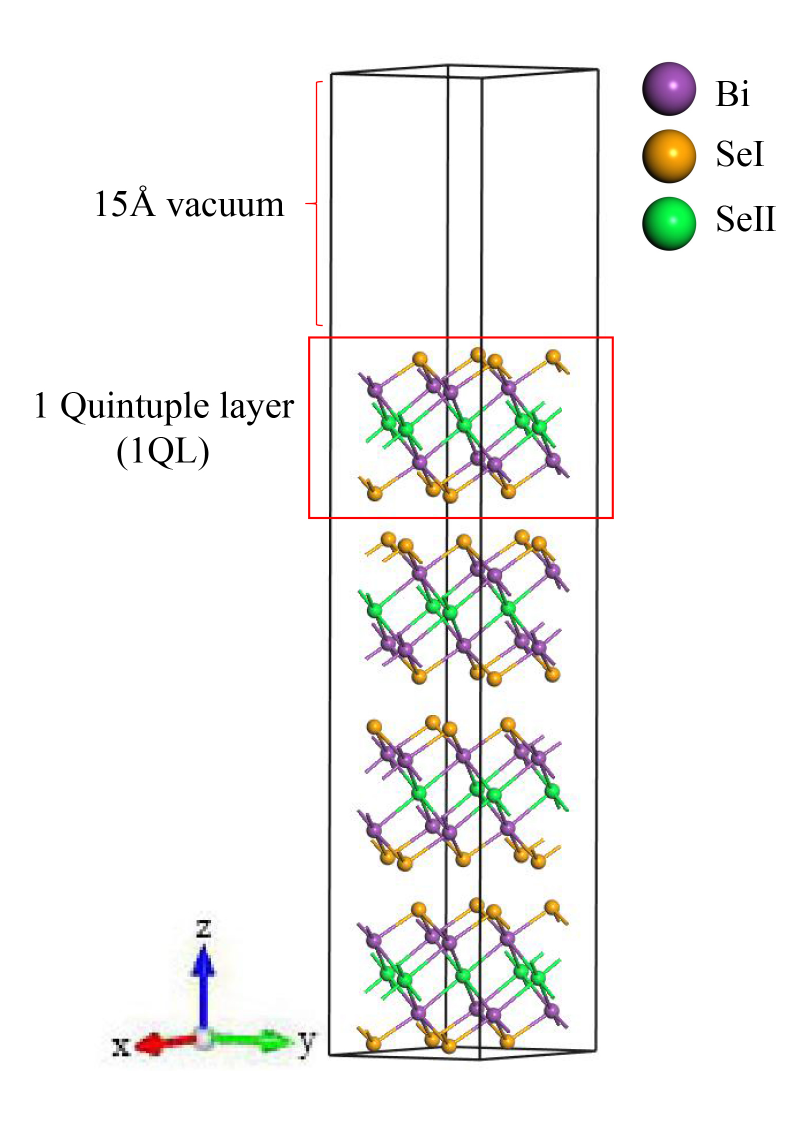}
    \end{center}
  \caption{\label{figm1} Ball and stick model of the Bi$_2$Se$_3$ surface.
    A four-quintuple-layer repeated 
    slab model is shown with 2$\times$2 surface cell, as 
    used in the adsorption studies.}
\end{figure}
\section{Computational details}
We performed DFT calculations
with the PBE exchange correlation potential and the projector
augmented wave method VASP~\cite{17,18,19}.
Van-der-Waals interactions were taken into account in the Grimme D-2 scheme.
The plane-wave cut-off energy was set at 500 eV.
The optimized lattice constants are $a$=4.12\AA\ and $c$=28.89\AA\ in very
good agreement with experiment ~\cite{20,21,22}. ($a$=4.137\AA\ and $c$=28.679\AA).
The Bi$_2$Se$_3$ surface was modeled with slabs of four QLs and
repeated slabs are separated by 15~\AA\ vacuum (Fig.~1).
The results were converged
with slab thickness as was checked with a few six QL calculations
(see Fig.~2 and Fig.~S2 in the S.I.).
The Brillouin zone of the 1$\times$1 surface is sampled with a
$\Gamma$-centered 8$\times$8$\times$1 k-point mesh.
Adsorption studies were performed with a 2$\times$2 surface supercell
and a 4$\times$4$\times$1 k-point mesh. The atomic positions of the adsorbate,
surface (Se) and subsurface layer (Bi) were fully optimized until
atomic forces were below 0.01~eV/\AA.
Atoms in deeper layers were fixed to their bulk positions.
Spin-orbit coupling was neglected during structural optimization.
Finally, the total energy, band structure and Fermi energy were
recalculated self-consistently including spin-orbit coupling,
with four times as many k-points, i.e. on a 8$\times$8$\times$1 mesh for
the 2$\times$2 surface cell.
\section{Results and Discussion}\label{sec3}
\begin{figure}
  \begin{center}
    \includegraphics[width=\columnwidth]{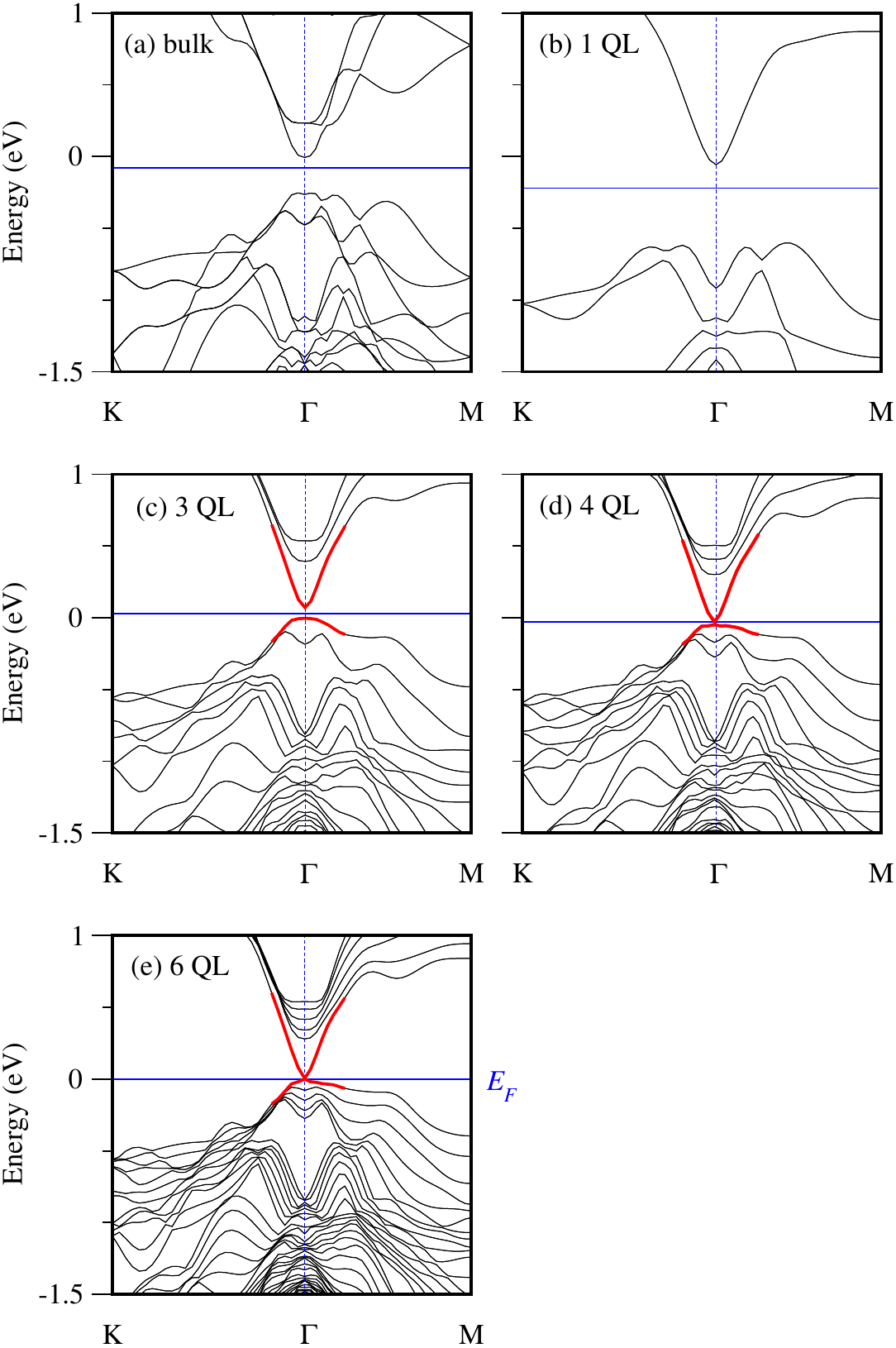}
    \end{center}
  \caption{\label{fig1} Calculated band structure of Bi$_2$Se$_3$
    along M-$\Gamma$-K for (a) bulk and (b-e) slabs of 1, 3, 4 and 6
    quintuple layers (QL).
    The Fermi energy is indicated by the blue, solid line and
    $E=0$ is defined as the Fermi energy of the clean 6~QL slab (e).
    The red colored bands indicate the Dirac cone.}
\end{figure}
\subsection{Clean surface and band alignment}
In Fig.~2~a--d the computed band structure along the high-symmetry line
M-$\Gamma$-K of the surface Brillouin zone is shown for bulk
Bi$_2$Se$_3$ and for slabs of different thicknesses.
It can be seen that the topological bands in the bulk band gap
($-0.2$~eV~$< E <0.1$~eV) develop progressively with increasing slab
thickness, and for a thickness of four or more QL the Dirac cone appears
at the Fermi level in agreement with the literature~\cite{23}.

We now turn to adsorption of chemical species that were 
present in the experiments of Ref.~\cite{16},
namely water, hydroxyl groups, oxygen and carbon.
We examine whether these different adsorbates can give rise to surface doping
with n- or p-type carriers and a corresponding
up- or down-shift of the Fermi level w.r.t. the clean surface
band structure.
We set the Fermi level of the clean surface (6 QL slab) to zero.
In all other systems, the energy scale is fixed by aligning
the lowest valence band of the adsorbed system with that of the
clean, 6-QL slab, see Fig.~S1 in the S.I. for details.
In this way, by fixing the energy zero to the Fermi-level of the clean surface,
the surface doping can be seen directly as a shift of the Fermi level,
which is indicated as a red line in the band structure plots in Figs~4,6.

\subsection{Adsorption of H$_2$O, OH, O, C and CH on the pristine surface}

\begin{figure}
  \begin{center}
  \includegraphics[width=0.8\columnwidth]{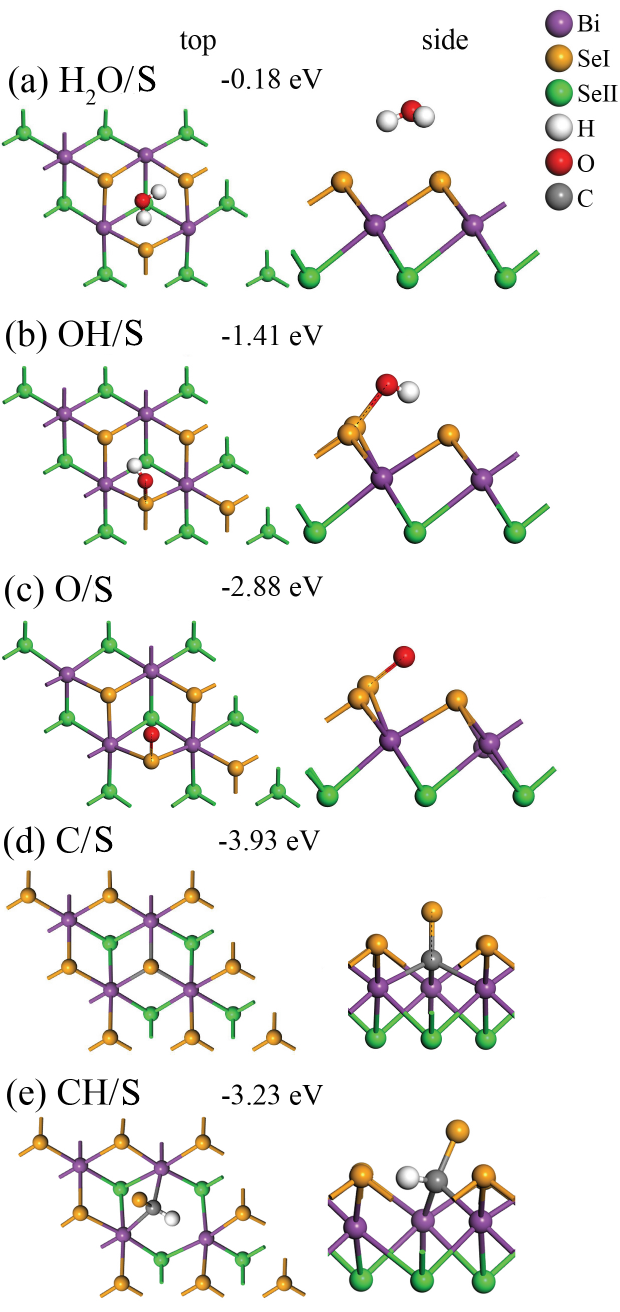}
    \end{center}
  \caption{\label{figstrnoV} Ball-and-stick models of
    most stable structures and adsorption energies
    $E_{\rm ads}$ of H$_2$O, OH, O and C adsorbed on a defect-free
    Bi$_2$Se$_3$ surface (denoted S)
    with a coverage of one molecule per 2$\times$2~cell.
  Top (left) and side (right) views are shown.}
\end{figure}
We consider a coverage of one adsorbed molecule (or atom) per 2$\times$2 surface
cell which has an area of 59~\AA$^2$.
We first study the adsorption of H$_2$O, OH, O, C and CH on the defect-free
Bi$_2$Se$_3$ surface.
The adsorption energy is defined as
\begin{equation}\label{eads}
E_{\rm ads} = E(X/S) - E(S) - E(X)
\end{equation}
where $E(X)$, $E(S)$ and $E(X/S)$ denote, respectively, the energy of specie X in
the gas phase, the energy of the clean surface (S) and the energy of X adsorbed on the surface.
For each adsorbate, several stable structures were found in the calculation.
Fig.~3 shows the most stable ones and their adsorption energies.
The H$_2$O molecule (Fig.~\ref{figstrnoV}~a) physisorbs on the Bi$_2$Se$_3$
surface with an adsorption energy of $-0.18$~eV, in qualitative agreement
with Ref.~\cite{5}.
The water molecule is located above the center of three Se surface atoms (SeI)
at a Se-O distance of 3.35~\AA\
and an orientation that hints to some weak Se-H interaction.
The hydroxyl (OH) molecule (Fig.~\ref{figstrnoV}~b) has a much larger
adsorption energy of $-1.41$~eV. The O atom makes
a covalent bond (bond length 1.89~\AA) with one Se surface atom. 
The Se-O-H angle is strongly bent,
with the H atom pointing to another surface Se atom
at a distance 2.59~\AA , indicating weak hydrogen bonding.
When a single O atom is adsorbed (Fig.~\ref{figstrnoV}~c) the O atom locates
at a similar position as in the OH case,
but the Se-O bond is shorter (1.69~\AA) and the
adsorption energy is twice as large ($-2.88$~eV),
as may be expected for a Se-O double bond.
A single C atom (Fig.~\ref{figstrnoV}~d) makes a strong reaction
with the surface, such that the Se atom is effectively replaced by the
C atom which bonds to Bi, while the Se atom is lifted and only bonds
to C with a bond length of 1.81~\AA. Note that the C atom has a
tetrahedral coordination with four covalent bonds, which
explains the high stability of the structure.
In the stable conformation of the CH adsorption (Fig.~\ref{figstrnoV}~e)
the C atom makes bonds to H, the lifted SeI and two Bi atoms.

\begin{figure}
  \begin{center}
   \includegraphics[width=\columnwidth]{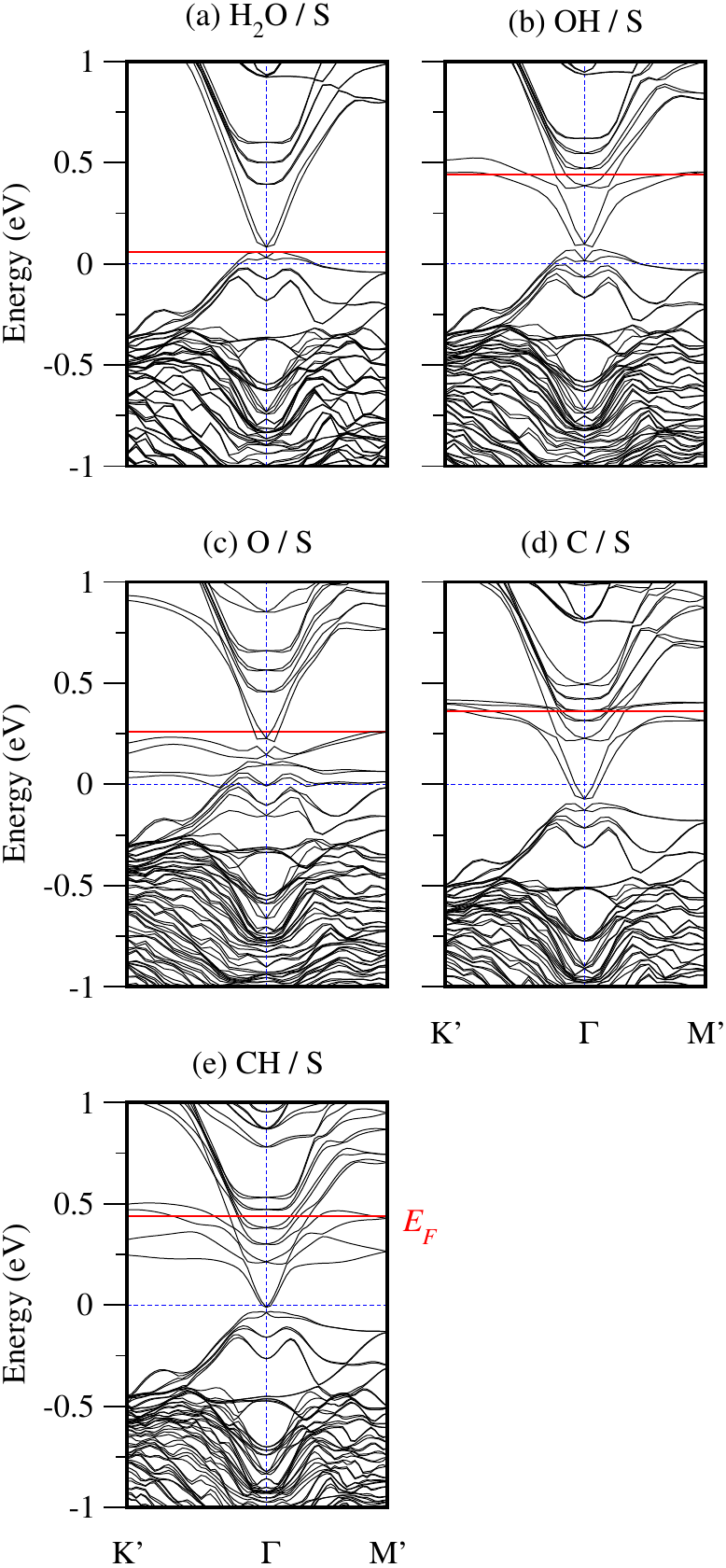}
    \end{center}
  \caption{\label{figbndnoV} Band dispersion of the adsorption structures
    on the defect-free surface (S) shown in Fig.~\ref{figstrnoV}.
    High-symmetry points of the 2$\times$2 surface 
    Brillouin zone are denoted by K' and M' and correspond to $k$-vectors at the midpoint
    of the $\Gamma$-K and $\Gamma$-M line, respectively,
    in the primitive Brillouin zone of Fig.~2. The red solid line shows the Fermi level
    and the blue dotted lines are a guide for the eye ($k$=0,$E$=0).}
\end{figure}

The electronic bands of these adsorbate structures are shown in Fig.~4.
The Brillouin zone of the 2$\times$2 cell is smaller than that of the 1$\times$1 cell,
and the bands are downfolded. M’ and K’ denote the M and K points of the 2$\times$2
cell and they correspond to the mid-points of $\Gamma$-$M$ and $\Gamma$-$K$ lines,
respectively, in the 1$\times$1 Brillouin zone.
When comparing Fig.~4~a and Fig.~2~d, it is seen
that water adsorption hardly changes the band
structure around the Fermi level, which is slightly shifted upwards by
$0.06$~eV, but still coincides with the Dirac point.
In other words, water adsorption has virtually no doping effect on clean
Bi$_2$Se$_3$, as expected from the fact that water is physisorbed.
Adsorption of an OH group (Fig.~4~b) however, leads to a large n-doping
with the Fermi-level moving up by $+0.44$~eV.
Also adsorption of atomic oxygen (Fig.~4~c) or atomic carbon (Fig.~4~d)
leads to a substantial n-doping with a Fermi
level shifts of $+0.26$~eV and $+0.36$~eV respectively.
The adsorption energies and Fermi energies are summarized in Table~1.
In conclusion, all considered species except H$_2$O, 
if adsorbed on the pristine surface would lead to n-doping.
\begin{figure}[h]
  \begin{center}
    \includegraphics[width=0.8\columnwidth]{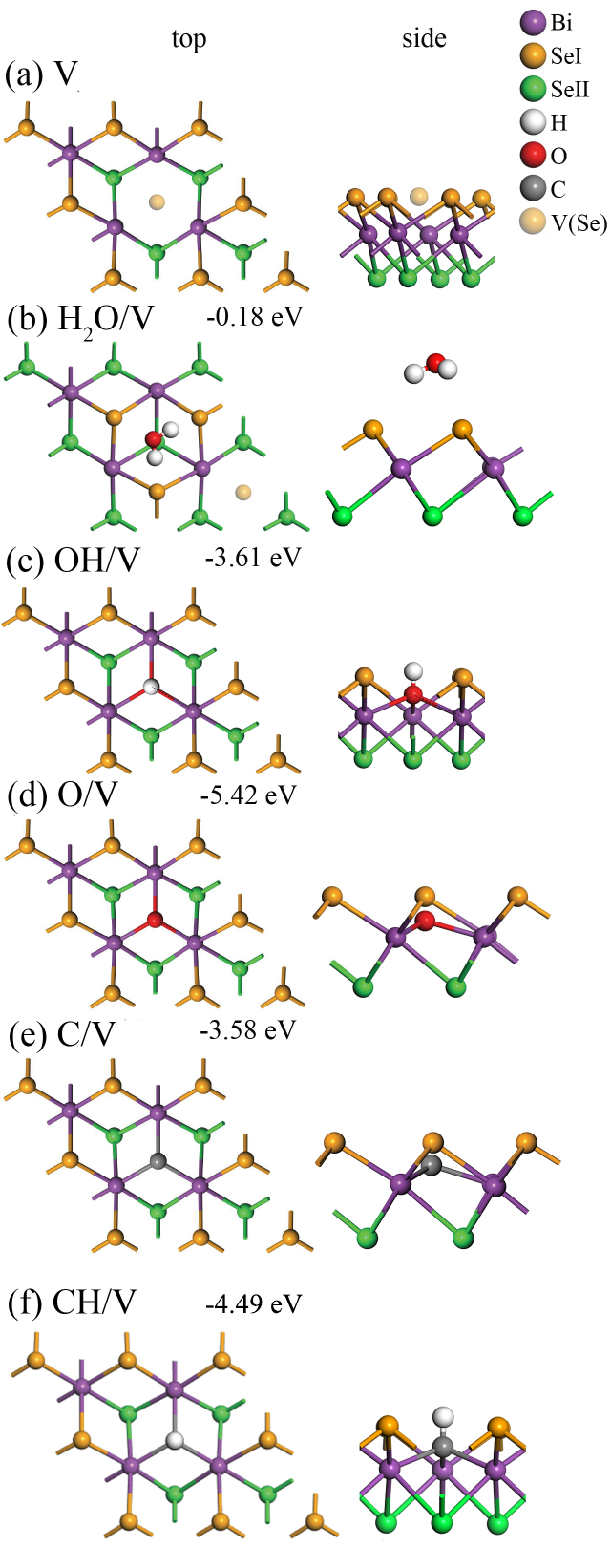}
    \end{center}
  \caption{\label{figstrV} (a) Computed structure of the Bi$_2$Se$_3$ surface
    with one SeI vacancy, whose position is indicated as a pale yellow ball (V(Se)).
    (b-e) Most stable structures and adsorption energies
    $E_{\rm ads}$ of H$_2$O, OH, O and C adsorbed on 
      the Se defective surface (which is denoted ``V''). 
  }
\end{figure}

\begin{table}
  \begin{tabular}{|l|r|r|r|r|}
    \hline
     &\multicolumn{2}{|c|}{X/S}  & \multicolumn{2}{c|}{X/V} \\
    \hline
X & \multicolumn{1}{|c|}{$E_{\rm ads}$} & \multicolumn{1}{c|}{$E_F$}
    & \multicolumn{1}{c|}{$E_{\rm ads}$} & \multicolumn{1}{c|}{$E_F$}\\
    \hline
H$_2$O & -0.18 & 0.06 & -0.18 & 0.29 \\ 
OH & -1.41 & 0.44 & -3.61 & 0.55 \\ 
O & -2.88 & 0.26 & -5.42 & 0.00 \\ 
C & -3.93 & 0.36 & -3.58 & 0.07 \\ 
CH & -3.23 & 0.44 & -4.49 & 0.20 \\
\hline
  \end{tabular}
  \caption{Adsorption energy $E_{\rm ads}$
    and Fermi energy $E_F$ (both in eV) of various species X adsorbed on the
    pristine Bi$_2$Se$_3$ surface (X/S) or on the Se defective surface (X/V).
  The Fermi level of the clean defective surface (V) is 0.32~eV}
\end{table}

\subsection{Adsorption on the Se defective surface}
Now we consider a defective Bi$_2$Se$_3$ surface with one Se vacancy in the surface
layer (SeI site) of the 2$\times$2 surface cell.
If the Se vacancy is located in the sub-surface layer (SeII site),
it is less stable by 0.52~eV. 
The optimized structure of the surface with one Se vacancy (``V(Se)'') is shown
in Fig.~5~a. This SE defective surface is denoted ``V'' in the following
and its band structure is shown in Fig.~6~a.
When comparing with the clean surface (Fig.~2~d),
it is seen that the band structure is very similar,
except for the appearance of a new band in $0<E<0.3$~eV
(which appears as two bands in the 2$\times$2 down-folded Brillouin zone).
The Dirac cone is still clearly visible at $\Gamma$ and $E\approx 0$.
However, the Fermi level is shifted upward by $+0.33$~eV
and thus reaches the bottom of the conduction band, which implies
that the carrier states are no longer topological edge states.
These findings agree with the literature~\cite{6,12,16}.
\begin{figure}[h]
  \begin{center}
    \includegraphics[width=\columnwidth]{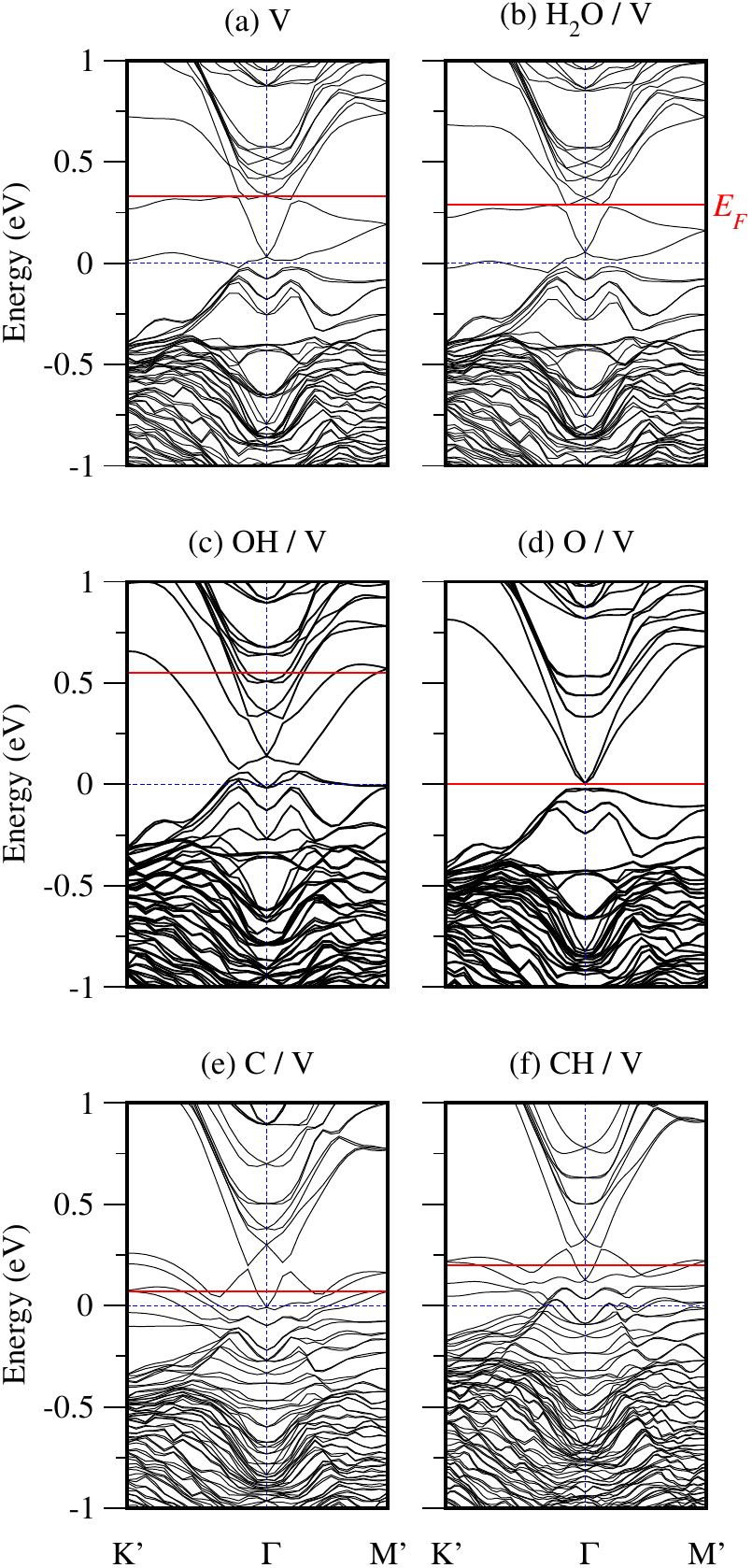}
    \end{center}
  \caption{\label{figbndV} Band dispersion of the adsorption structures
    of Fig.~\ref{figstrV} on the Se-defective surface (V)}
\end{figure}
Next, we have studied the adsorption of H$_2$O, OH, O, C and CH
on the Se-defective surface. 
For each adsorbate X, the most stable structure, denoted X/V, 
is shown in Figs~5~b--d and the adsorption energies are indicated. They are
calculated using Eq.~(\ref{eads}) upon replacing S by V.

The water molecule is adsorbed at the same position and orientation as in
the defect-free case (Fig.~3~a), i.e. above the center of three SeI surface
atoms, and the adsorption energy is nearly the same ($-0.18$~eV).
The band structure of the defective surface (Fig.~6~a) is almost unchanged
upon water adsorption and the Fermi level is at the same position as before
adsorption.
When a hydroxyl group is adsorbed on the defective surface
the OH molecule fills the Se vacancy site (Fig.~5~c).
The O atom substitutes the missing Se atom and makes bonds with the three neighboring Bi atoms (bond length 2.55\AA).
The OH adsorption energy ($-3.65$~eV) is over 2~eV larger
than on the pristine surface. 
The band structure is shown in Fig.~6~c. As compared to the clean
surface, the Dirac cone is slightly shifted (by $+0.15$~eV)
but still clearly visible.
The Fermi level is at $E_F=+0.55$~eV, which is the highest value found for all
systems studied here. As compared to the clean, defective surface
(Fig.~6~a, $E_F=0.32$~eV), the Fermi level is further up-shifted
by OH-adsorption. So the OH-species always leads to n-doping, whether it
is adsorbed on the pristine surface (Fig.~3~b) or on a Se-vacancy site.
Among the species considered here, this universal n-doping behaviour is only
found for the OH-group. 
In the experiments~\cite{16} it was found that water adsorption followed
by XUV irradiation can, under certain conditions,
lead to n-doping with a Fermi-level shift of about 0.1~eV.
At the same time, the O-1s XPS data showed a peak with a binding energy
between that of H$_2$O and most oxides (O$^{2-}$) which may be due to a
hydroxyl group.
Therefore, the present findings strongly suggest that the experimentally
observed n-doping is due to OH-species, produced by radiative dissociation
of water molecules.

When a single O atom is adsorbed on the defective surface (Figs.~5~d)
the O atom occupies the Se vacancy site and makes covalent bonds with
the three neighboring Bi atoms, as in the case of OH adsorption.
The adsorption energy of O/V is $-5.42$~eV, i.e. about twice larger
than that of O on the pristine surface (O/S).
Interestingly, the band structure of O/V system is very similar to that
of the pristine, defect-free surface (Fig.~2~d).
The Fermi-level is exactly at the Dirac point ($E_F$=0) and so
the surface is neither n- nor p-doped.
This is because oxygen is isoelectronic to selenium, 
and one O atom can passivate almost perfectly the dangling bonds created by one Se
vacancy. We thus find that oxygen adsorption on the defective surface gives rise to
p-doping which can completely undo the n-doping effect of the Se vacancies,
in agreement with Koleini et al.~\cite{12}.
When carbon is adsorbed on the defective surface, the C atom also fills
the Se vacancy and the Fermi level is down-shifted as compared to the clean defective surface by about 0.2~eV.
The $p$-doping effect is, however smaller than for the
O atom, and the band structure at $E\approx 0$ around the $\Gamma$-point
is different from the pristine surface, with several new bands appearing
near the Dirac cone.
The same is true for CH adsorption on the Se vacancy.

For the discussion of possible surface reactions, we have
listed the formation energies of the various adsorbate systems in Table~1,
calculated with the following standard states:
atomic C, molecular H$_2$ and O$_2$ as well as the clean Bi$_2$Se$_3$ surface
either pristine or with one Se vacancy.
\begin{table}[!ht]
    \centering
    \begin{tabular}{c|cccccc}
      X     &    H  &  O    &  C    &  OH   &   CH  & H$_2$O \\
      \hline
E(X/S)&  0.938 & -2.574 & -3.973 & -3.704 & -4.630 & -5.728 \\
E(X/V)&  0.502 & -5.120 & -3.576 & -5.902 & -5.892 & -5.722 \\
   \end{tabular}
    \caption{Calculated formation energy $E$ (in eV)
      of various chemical species
      X adsorbed
      on the pristine (S) and Se-defective (V) Bi$_2$Se$_3$ surface.}
\end{table}
From Table 1, various reactions energies can be easily computed, e.g.
for the reaction
H/S~+~CH/S $\rightarrow$ C/S~+~H$_2$, we have
$\Delta_rE = E$(C/S)$-E$(CH/S)$-E$(H/S)$= -0.28$~eV.

\subsection{Analysis of experimental data in the literature}
A major aim of this study is to shed light on the
experimental results of Ref.~\cite{16},
which can be summarized as follows.

(i) The clean Bi$_2$Se$_3$ surface is $n$-doped due to intrinsic defects, 
mainly Se vacancies. The Fermi-level is about 0.4~eV above the Dirac point.

(ii) Water adsorption together with UV- or soft X-ray irradiation
leads to $p$-doping, i.e. a down shift of the Fermi-level
of up to 0.3~eV. As a result, the Dirac point is only about 0.1~eV
below the Fermi level and the topological nature of the surface electronic
structure is restored.
Apart from the shift of the Dirac point, no changes of the
band structure were detected.

(iii) $p$-doping was only observed when the surface is
contaminated with a small amount of carbon, which comes from the
scotch tape used for cleaving. For a carbon-free surface, water adsorption
and light irradiation was found to slightly increase the $n$-doping.

(iv) The O-1s core-level spectra of H$_2$O covered surface is at a
binding energy $E_B=533$~eV. Upon light radiation, two new peaks appear at
lower binding energy, namely $E_B=531$~eV and $E_B=529.5$~eV.
The $E_B=529.5$~eV peak systematically appeared when the surface
became $p$-doped.
The peak at $E_B=531$~eV, however, was not always observed,
suggesting that the peak may be the signature of an intermediate state in the
$p$-doping process.

In section~\ref{sec3} we have examined the band structure changes
induced by adsorption of H$_2$O, OH, O, C and CH
on the pristine and Se defective surfaces.
In experiment the p-doping manifested itself as a shift of the Dirac point,
without other changes of the band structure.
H$_2$O molecules physisorb on the surface and have almost no doping effect.
OH species can fill the Se vacancies, but this leads to
further n-doping, rather than p-doping.
Adsorption of C at the Se vacancy leads to p-doping, as
seen in Fig.~6~e, but the band structure around the Dirac point is strongly
modified with the appearance of new, topologically trivial bands.
For a high C impurity concentration as in the present model calculation,
these bands would destroy the topological nature of the surface states.
In the experiments of Ref.~\cite{16} no such band structure changes were observed.
When an oxygen atom is adsorbed at the Se vacancy,
we find strong $p$-doping which leaves the Dirac cone unchanged
and totally removes the Se-vacancy induced $n$-doping, in
agreement with experiment.
Therefore, among the species considered, only oxygen adsorption at
the Se vacancies can account for the experimental findings of Ref~\cite{16}.
\begin{figure}[h]
  \begin{center}
   \includegraphics[width=\columnwidth]{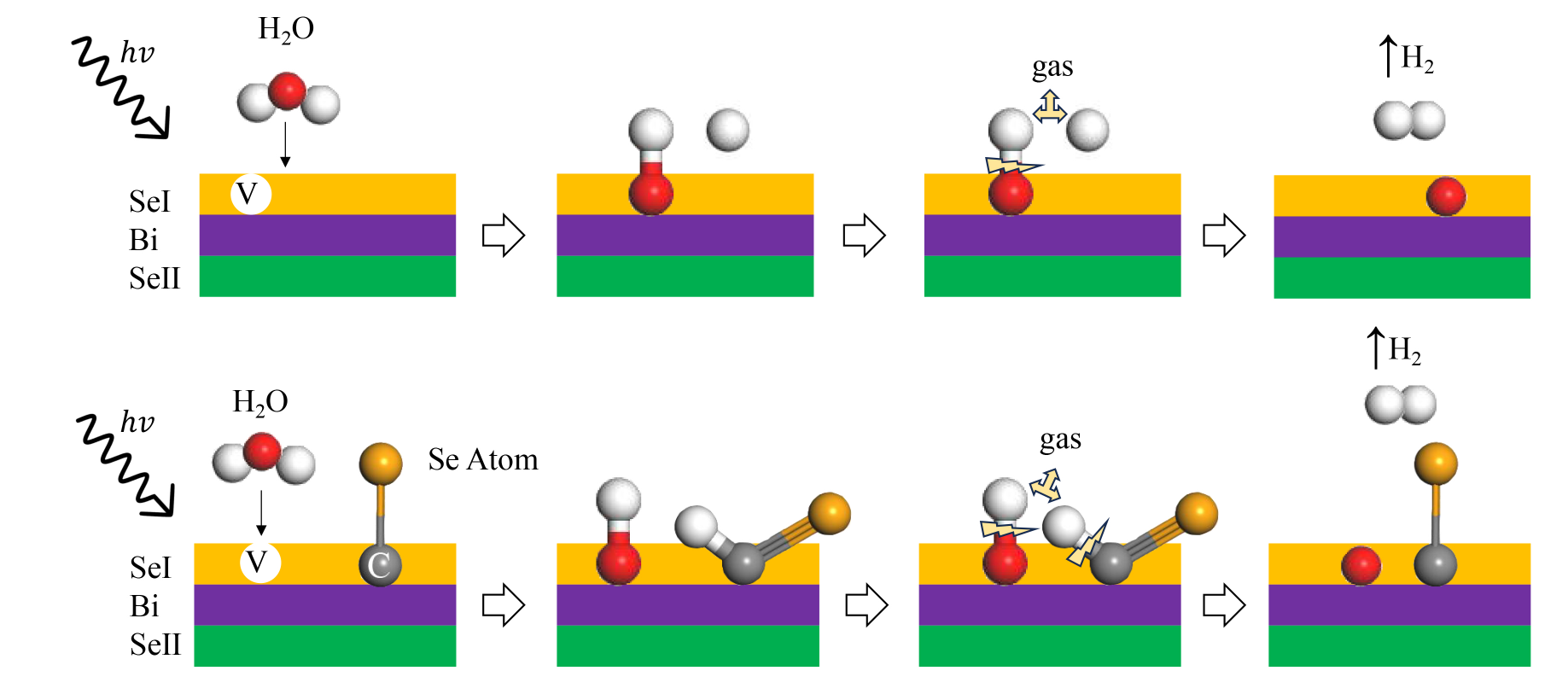}
    \end{center}
  \caption{\label{figm7} Scheme of the suggested reaction mechanism for
    the XUV-induced $p$-doping of a Se-defective Bi$_2$Se$_3$ surface
    without (upper panel) and with (lower panel) C impurities.}
\end{figure}
As there was no free oxygen present in the experiment~\cite{16}
the adsorbed O atoms must have been provided by water splitting
according to the following total reaction
\begin{equation}\label{vh2o2vo}
  {\rm H}_2{\rm O}/{\rm V} \rightarrow {\rm O}/{\rm V} +{\rm H}_2
  \quad \Delta_rE = 0.60{\ \rm eV} 
\end{equation}
where
$\Delta_rE$ is the calculated reaction energy.
The reaction is endothermic and needs an external supply of energy.
This is in agreement with the experiment. Indeed, Bi$_2$Se$_3$ is chemically
very stable~\cite{5}. The water covered Bi$_2$Se$_3$ shows a chemical reaction
only upon intense UV or X-ray irradiation~\cite{16},
where the adsorption of photons and the following
hole-decay processes can provide the energy needed to trigger an endothermic
surface reaction. It is clear that once molecular hydrogen is formed, it will
immediately desorb from the surface for entropy reasons.
Therefore reaction (\ref{vh2o2vo}) is in practice irreversible,
which agrees with irreversibility of the $p$-doping process in experiment~\cite{16}.
The foregoing analysis shows that filling of Se vacancies by oxygen from water
molecules is possible on thermodynamic grounds and that it provides the only
explanation for the observed $p$-doping which fits the experimental data~\cite{16}.

\subsection{A model for the chemical reaction mechanism}
The atomistic details of the reaction are, however, far from obvious.
Water splitting at a solid surface generally involves several intermediate
states~\cite{24}. It is highly unlikely that reaction~(\ref{vh2o2vo})
occurs directly in a single step,
because this would require the simultaneous breaking of two O-H bonds.
Based on the theoretically calculated energies in Table~I,
we propose a more likely reaction mechanism, which
is schematized in Fig.~\ref{figm7} and explained in the following.
We start from a water molecule physisorbed at the Se vacancy.
By photon adsorption, enough energy is provided to break one O-H and release
a H atom according to the reaction
\begin{equation}\label{vh2o2voh}
  {\rm H}_2{\rm O}/{\rm V} \rightarrow {\rm OH}/{\rm V} +{\rm H}/{\rm S}
  \quad \Delta_rE = 0.76{\ \rm eV} 
\end{equation}
While the reaction energy is slightly higher than that of~(\ref{vh2o2vo})
it is much more likely since it involves only one O-H bond breaking.
Once the H atom is released, it will diffuse on the Bi$_2$Se$_3$ surface.
It may go back to the vacancy site in which case the reverse
reaction of~(\ref{vh2o2voh}) occurs and the H$_2$O molecule is restored.
This is most likely, but there is a non-zero chance for two competing reactions.
First, it is possible that two diffusing H atoms meet and form molecular
hydrogen:
\begin{equation}\label{hshs_h2}
  {\rm H}/{\rm S} + {\rm H}/{\rm S} \rightarrow {\rm H}_2
  \quad \Delta_rE = -1.88{\ \rm eV} 
\end{equation}
Second, when carbon impurities are present,
the diffusing H atoms may be trapped
at the C sites according to the reaction
\begin{equation}\label{hscs_chs}
  {\rm H}/{\rm S} + {\rm C}/{\rm S} \rightarrow {\rm CH}/{\rm S}
  \quad \Delta_rE = -1.60{\ \rm eV} 
\end{equation}
The reaction energy was calculated with the most stable conformations for
C/V and CH/V, see Figs~3~d,e.
Given the large exothermic reaction energy, the trapped H atom is very stable.
When a second diffusion H atom reaches the C site, molecular hydrogen can be
formed as
\begin{equation}\label{hschs_csh2}
  {\rm H}/{\rm S} + {\rm CH}/{\rm S} \rightarrow {\rm C}/{\rm S} + {\rm H}_2
  \quad \Delta_rE = -0.28{\ \rm eV} 
\end{equation}
which is also exothermic.
Reactions (\ref{hscs_chs}) and (\ref{hschs_csh2}) together yield the same
product (molecular hydrogen) as reaction (\ref{hshs_h2}), i.e.
the carbon impurities acts as a catalyst.
Whether the hydrogen formation occurs directly (\ref{hshs_h2}) or
with the C catalyst, depends on the relative concentrations of adsorbed
atomic hydrogen and of C impurities.
In the experiments~\cite{16} the C concentration was in the 0.1--1\%
range. The concentration of adsorbed H atoms produced by the photo-assisted
reaction~(\ref{vh2o2voh}) is very difficult to estimate. However,
given the large endothermic reaction energy, it might be well below 0.1\%.
In this case, the hydrogen formation reaction would be driven by
the C catalyst.
Once a Se vacancy is filled by a hydroxyl group, the following reaction may
occur
\begin{equation}\label{ohv_ov_hs}
  {\rm OH}/{\rm V} \rightarrow {\rm O}/{\rm V} + {\rm H}/{\rm S} 
  \quad \Delta_rE = 1.72{\ \rm eV} 
\end{equation}
This reaction is clearly endothermic which means that OH/V is a stable
intermediate state of the water splitting reaction.
Under XUV irradiation, photon absorption may provide the
energy $\Delta_rE$ for reaction~(\ref{ohv_ov_hs}). Moreover, hydrogen
dissociation may be facilitated by the fact that the photoemission process
leads to a positive charging of the surface~\cite{16}. 
Indeed, if the photohole is located on the OH group, a
H$^+$ ion may detach and form H$_3$O$^+$ with a nearby water molecule.
Since reaction (\ref{ohv_ov_hs}) is by 1~eV more endothermic 
than hydrogen dissociation from water (reaction \ref{vh2o2voh}),
its reaction rate should be much smaller, which implies that the
concentration of H atoms at the surface ($c_{\rm H}$) is very low.
Then, direct H$_2$ formation by two diffusing H atoms
(reaction~\ref{hshs_h2}) seems very unlikely,
since the reaction rate is proportional $c_{\rm H}^2$.
The rate of reaction (\ref{hschs_csh2}), on the other hand, is proportional to
$c_{\rm H}c_{\rm C}$, where $c_{\rm C}$
is the concentration of C surface impurities.
When C impurities can be detected by XPS (as in Ref.~{15})
we can safely assume $c_{\rm C} \gg c_{\rm H}$, and so reaction (\ref{hschs_csh2})
is far more likely than~(\ref{hshs_h2}), i.e.\  H$_2$
formation is much facilitated by the presence of C impurities.
The C impurities form a reservoir for surface hydrogen
(reaction \ref{hscs_chs}) and thus 
can act as a catalyst for H$_2$ formation at the Bi$_2$Se$_3$ surface
(reaction \ref{hschs_csh2}).
This may explain the experimental finding (iii) above, i.e. the fact that
the XUV induced p-doping process
was only observed at surfaces with some carbon contamination~\cite{16}.

In order to better understand the chemical shifts observed at the O-1s
core-level (point (iv) above) we have computed the binding energies ($E_B$)
with DFT-VASP (see the S.I. for details).
The calculated binding energies decrease in the sequence
$E_B$(H$_2$O/S) $>$  $E_B$(OH/V) $>$ $E_B$(O/V).
By comparing the theoretical and experimental chemical shifts~\cite{16},
we can attribute
the peak of highest binding energy (533 eV),
which is the only peak before irradiation,
to adsorbed H$_2$O.
The low energy peak (529.5 eV) which appears concomitantly with p-doping
is attributed to O/V and the intermediate-energy peak (531 eV) to
OH/V. These assignments fully support the proposed reaction mechanism,
where OH/V is a quite stable intermediate state
and O/V is the final state of the water splitting reaction.

\section{Conclusions}
In summary, we have presented a first-principles study on the adsorption
of H$_2$O, OH, O, C and CH on a Bi$_2$Se$_3$ surface,
either pristine or with Se vacancies. The effect of the
adsorbed species on the topological insulator band structure was investigated.
We find that H$_2$O physisorbs on Bi$_2$Se$_3$ and has a negligible doping
effect. All other species interact with the surface by chemisorption.
On the pristine surface, adsorption of OH, C or O adsorption leads to
$n$-doping. The Se-defective surface is intrinsically $n$-doped.
If an OH group adsorbs at the Se-vacancy, the $n$-doping increases even more
but if a C or a O atom fills the vacancy, $p$-doping occurs.
In the case of C, the band structure around the Dirac
point is strongly changed, but in the case of O, the topological surface
band structure of pristine Bi$_2$Se$_3$ is perfectly restored.
We have discussed a recently reported method for controlled surface
$p$-doping which uses water adsorption and XUV irradiation~\cite{16}.
Based on the computed adsorption energies, we 
propose the following reaction mechanism which may explain
the experimental findings of Ref.~\cite{16}.
Water decomposes into an O atom which fills the Se vacancy and a hydrogen
molecule which desorbs from the surface. The reaction is
weakly endothermic but the required energy (0.6~eV) can be easily
provided by photon absorption. Since the simultaneous breaking of two OH
bonds is highly unlikely, we suggest that in a first step an OH group
is formed which fills the vacancy while the released H atom binds to Se.
In a second step the remaining H atom of the adsorbed OH group
is released under photo-absorption. This reaction is substantially 
more endothermic (by about 1 eV) than the first step and should 
thus be the rate limiting process. As a consequence, the adsorbed OH group
forms a relatively stable intermediate state during XUV irradiation.
The presence of an intermediate state was indeed evidenced as a peak
in the experimental O-1s XPS~\cite{16}
whose chemical shift qualitatively agrees with the theoretical
value of the adsorbed OH group.
Adsorbed C atoms can trap hydrogen atoms that are diffusing on the
surface and thus may act as a hydrogen atom reservoir for H$_2$ molecule
formation. In this way, adsorbed C atoms may play a catalytic
role in the light-induced water splitting reaction, in agreement
with experiment~\cite{16}.

\subsection*{Acknowledgments}
C.F. is grateful for financial support by JST SPRING, grant No. JPMJSP2109.
K.S. acknowledges support by JSPS KAKENHI grant No. JP22H01957
and JP20H05621 and the Spintronics Research Network of Japan.

\bibliographystyle{elsarticle-num}
\bibliography{cas-refs.bib}
\end{document}